\documentclass[twocolumn,amsmath,amssymb,nofootinbib,superscriptaddress]{revtex4}
\usepackage{setspace}
\usepackage{graphicx}
\usepackage{psfrag}
\usepackage[hyperindex,breaklinks]{hyperref}
\def\pmb#1{\setbox0=\hbox{$#1$}%
  \kern-.025em\copy0\kern-\wd0
  \kern.05em\copy0\kern-\wd0
  \kern-.025em\raise.0433em\box0}
\def\parb{\pmb{\partial}}
\def\alt{\mathrel{\hbox{\rlap{\hbox{\lower4pt\hbox{$\sim$}}}\hbox{$<$}
}}}

\def\be{\begin{equation}}
\def\ee{\end{equation}  }
\def\bea{\begin{eqnarray}}
\def\eea{\end{eqnarray}  }

\begin{document}
\title{Evolution to a smooth universe in an ekpyrotic contracting 
phase with $w 
> 1$}

\author{David Garfinkle}
\email{garfinkl@oakland.edu}
\affiliation{Department of Physics, Oakland University, Rochester, MI 48309}
\author{Woei Chet Lim}
\email{wlim@princeton.edu}
\affiliation{Joseph Henry Laboratories, Princeton University, Princeton, NJ 08544}
\author{Frans Pretorius}
\email{fpretori@princeton.edu}
\affiliation{Joseph Henry Laboratories, Princeton University, Princeton, NJ 08544}
\author{Paul J. Steinhardt}
\email{steinh@princeton.edu}
\affiliation{Joseph Henry Laboratories, Princeton University, Princeton, NJ 08544}
\affiliation{Princeton Center for Theoretical Science, Princeton University, Princeton, NJ 08544 }

\begin{abstract}
A period of slow contraction with equation of state 
$w > 1$, known as an ekpyrotic phase, has been shown to flatten and 
smooth  the universe if it begins the phase with small perturbations.  
In this paper, we explore how robust and powerful the ekpyrotic 
smoothing mechanism is by beginning with 
highly inhomogeneous and anisotropic initial conditions
and numerically solving for the subsequent evolution of the universe.  
Our studies,  based on a universe with gravity 
plus  a scalar field with a negative exponential potential,
show  that some regions become homogeneous and isotropic while others 
exhibit inhomogeneous and anisotropic behavior in which the scalar field behaves
like a fluid with $w=1$.  We find that the ekpyrotic smoothing 
mechanism is robust in the sense that the ratio
of the proper volume of the smooth to non-smooth region
grows  exponentially  fast along time slices of constant mean 
curvature.
\end{abstract}

\maketitle

\section{Introduction}\label{sec_intro}

For over two decades, the only known mechanism for homogenizing, 
isotropizing and flattening the universe was inflation, a period 
of accelerated expansion with an equation of state $w$
(ratio of  pressure to  energy density)  near -1.
Its success in resolving the horizon and flatness problems is a
principal reason why inflation became an essential part of the 
standard model of cosmology.  In recent years, an 
alternative mechanism has been discovered in which smoothing and 
flattening 
occurs before the big bang as the universe undergoes a 
period of slow contraction with $w >1$.  
This alternative, known as the ekpyrotic mechanism 
\cite{Khoury:2001wf,Erickson:2003zm}, has been incorporated in 
alternatives to standard big bang inflationary cosmology including the 
``ekpyrotic" \cite{Khoury:2001wf},  ``new ekpyrotic" 
\cite{Buchbinder:2007ad} 
and cyclic models \cite{Steinhardt:2002ih}.  We note that both the 
inflationary and ekpyrotic mechanisms can 
 produce nearly scale-invariant spectra for density perturbations in 
addition to smoothing and flattening the universe.

Until now, the ekpyrotic mechanism has only been shown 
to work in cases where the deviations from smoothness and 
flatness are small and perturbative when the ekpyrotic phase begins.  
The purpose of this paper is to show that the 
ekpyrotic mechanism is powerful and robust enough to smooth the 
universe even when the initial perturbations are large and non-linear.

Let us first review how the inflation and the ekpyrotic 
mechanism work in the perturbative regime where the cosmic evolution 
is well approximated  by the Friedmann equation with an anisotropy 
term:
\begin{equation} \label{eq1}
 H^2 = \frac{8 \pi G}{3} \left(\frac{\rho_m^0}{a^3} + 
\frac{\rho_r^0}{a^4}+
\frac{\rho_w^0}{a^{3(1+w)}} \right) - \frac{k}{a^2} + 
\frac{\sigma^2}{a^6}, 
\end{equation}
where $H \equiv \dot{a}/{a}$ is the Hubble parameter; $a(t)$ is 
the Friedmann-Robertson-Walker scale factor normalized so 
that the  value today, $t_0$, is $a(t_0)=1$; 
$\rho_i^0$ represents the present 
value of the energy density for component $i$, where $m$ represents 
non-relativistic matter, $r$ represents radiation and $w$ 
represents an energy component with equation of state $w$, 
such as a scalar field and its potential. The 
last two terms on the right-hand side represent 
spatial curvature and anisotropy.  

For an expanding universe, the term that dominates Eq.~(\ref{eq1}) 
after a long period of expansion is the one with 
the smallest power of $a$ in the denominator.  With 
only radiation and matter, the dominant term would be the spatial 
curvature, leading to a universe that is unacceptably open or closed 
by the present epoch.  However, introducing an energy component with 
$w \approx -1$ totally changes the outcome  because this component 
($\rho_w$)
then has the smallest exponent and dominates the Einstein equation, 
while 
the curvature and anisotropy (and other energy components) become 
negligible.
This is the essence of how inflation works if the initial conditions 
are perturbative.  For inflation, there are several
``Cosmic No-hair'' theorems in addition to numerical
results supporting the claim that homogeneity, isotropy and flatness 
develop even 
when the initial conditions are non-linear and 
non-perturbative~\cite{kolb_turner_86,Goldwirth}.

Now consider the analogous arguments for a contracting universe.
With $a(t)$ shrinking, the dominant term in Eq.~(1) will  
be the one with the largest exponent of $a$ in the 
denominator.  For a universe with matter and radiation only, this 
would be the anisotropy term, which famously overtakes the evolution 
and 
drives the universe into chaotic mixmaster behavior.  On the other 
hand, 
if there is an energy component with $w >+1$, then this energy 
component dominates instead  of anisotropy or spatial curvature, and 
chaotic mixmaster behavior never begins \cite{Erickson:2003zm}.  For 
a scalar field $\phi$ with potential energy density $V(\phi)$, the 
ratio of pressure to energy density is
\begin{equation}\label{eq2}
w \equiv \frac{ \frac{1}{2} \dot{\phi}^2 -V}{\frac{1}{2} \dot{\phi}^2 
+V},
\end{equation}
which can be significantly greater than unity when $V$ is 
less than zero and non-negligible and which 
approaches $w=1$ if the scalar field kinetic energy dominates.  
The ekpyrotic phase in ekpyrotic and cyclic models
includes an effective scalar field 
component of this type.

For the cyclic model, the ekpyrotic phase is preceded by a 
period of dark energy domination and accelerated expansion which, if 
sustained 
long enough, would make the universe uniform and flat before the 
ekpyrotic contraction phase begins.  In this case, the perturbative 
argument 
above should be reliable and sufficient to conclude that the universe 
is 
smooth and flat as it approaches the big crunch.  However, 
in the ekpyrotic or new ekpyrotic models, generally, or in 
the cyclic model with a very short dark energy phase, the conditions 
at the beginning of the ekpyrotic phase are under less control.

This paper investigates the robustness of the ekpyrotic smoothing and 
flattening 
mechanism when the initial conditions are non-linear and non-
perturbative 
to determine how the situation compares with the perturbative case and 
with  
inflation. For this study we are
not concerned with any particular form for the
initial conditions that may be motivated by some specific model; 
rather, we 
would like to understand how a generic, highly inhomogeneous and 
anisotropic space-time
evolves under the influence of the proposed smoothing mechanism, 
modeled here
by a scalar field with a negative exponential potential.
Such negative exponential potentials arise naturally in supergravity
and in string theory.
Investigation of the proposed smoothing mechanism 
requires numerical solution of the coupled Einstein-scalar 
system of equations.  For simplicity, we restrict our studies to 
deviations from smoothness along a single spatial dimension.

We use an orthonormal frame representation
of the equations written in terms of Hubble-normalized, 
scale invariant variables~\cite{Uggla:2003fp} similar to that
described in \cite{Curtis:2005va}, though here coupling
to a scalar field instead of a fluid, and using
constant-mean-curvature (CMC) time slices. We discretize
the equations using second-order accurate finite difference 
techniques, and
solve them with a variant of the Berger and Oliger~\cite{BO} 
adaptive mesh refinement (AMR) algorithm for coupled 
elliptic-hyperbolic equations~\cite{Pretorius:2005ua}.
We find smooth regions that are scalar field dominated in which 
the scalar field (kinetic plus potential energy density) 
component behaves like a fluid with $w \gg 1$, and also
regions where the scalar field kinetic energy dominates over the 
potential 
energy and the scalar field behaves like a fluid with $w=1$.  
These latter regions remain inhomogeneous and anisotropic,
and throughout this paper we will refer to these parts
of the universe as the ``anisotropic regions''.
Note however that the anisotropic regions are neither anisotropy nor matter 
dominated because both the scalar field and the anisotropy of the metric
play important roles in the dynamics.
Futhermore, note that despite the fact that matter in the smooth regions behaves 
effectively like a fluid with $w \gg 1$ there is no issue of superluminal
propagation as might arise from an actual fluid with such an equation
of state.  This is because we are always solving the scalar wave 
equation with potential where disturbances always propagate within 
the light cone. 
In the anisotropic regions ``spikes'' also form, which are places where the fields change
on very small spatial scales, and are similar to regions with this 
property that have been observerd in numerical simulations of 
singularities in vacuum spacetimes\cite{dgspike}.
Despite the presence of the scalar field, the anisotropic regions exhibit 
dynamical behavior similar to chaotic 
mixmaster vacuum solutions, where there are a series
of relatively quick transitions between longer epochs
where the solution can be described by a $w=1$ Bianchi type I spacetime.
A difference here though is that there are only a {\em finite} number
of transistions, so the mixmaster behavior terminates after
several transitions. These dynamics are also known to occur
in spacetimes where the matter {\em is} a fluid with $w=1$~\cite{Curtis:2005va, Coley05}.
AMR is necessary to resolve the spiky features that form both
in the anisotropic regions and, in some instances,
briefly in what will eventually become smooth 
scalar field dominated regions,
and to resolve the almost domain wall-like transitions that develop
between the smooth and anisotropic regions. 

The outline for the rest of the paper is as follows.
In Sec. \ref{sec_methods} we describe the equations, initial 
conditions,
and numerical methods used to solve them. We present the results
in Sec. \ref{sec_results}. The primary conclusion is that a
scalar field with a potential inspired by cyclic models
is a remarkably powerful and robust smoothing mechanism during a
contracting phase of the universe, able to drive the spacetime
to homogeneity and isostropy even starting with
highly non-linear deviations from an FRW spacetime. 
Concluding remarks and a discussion of future work
is given in Sec. \ref{sec_conclusion}.

\section{the equations and solution method}\label{sec_methods}

The method we use to evolve the Einstein-scalar equations is
the scale invariant tetrad method
of Uggla et al.
\cite{Uggla:2003fp}.
We use this method with constant mean curvature slicing as is done
in the vacuum simulations of
\cite{dgaei} but with scalar field matter instead of vacuum.  Thus
our system can be thought of as the system of \cite{dgaei} but with
extra variables and equations describing the matter, and with extra
source terms for the influence of the matter on the metric evolution
equations.  More information on this type of method can be found in
\cite{Uggla:2003fp,dgaei}.   

The spacetime
is described in terms of a coordinate system ($t,{x^i}$) and a tetrad
(${{\bf e}_0},{{\bf e}_\alpha}$)  where both the spatial coordinate
index $i$ and
the spatial tetrad index $\alpha $ go from 1 to 3.  Choose
${\bf e}_0$ to be hypersurface orthogonal with the relation between
tetrad and coordinates of the form
${{\bf e}_0} = {N^{-1}}{\partial _t}$, and 
${{\bf e}_\alpha} =
{{e_\alpha }^i}{\partial _i},$
where $N$ is the lapse and the shift is chosen to be zero.
Choose the spatial frame $\{ {{\bf e}_\alpha} \}$ to be
Fermi propagated along the integral curves of ${\bf e}_0$.
The commutators of the tetrad components are decomposed as follows:
\begin{eqnarray}
[{{\bf e}_0},{{\bf e}_\alpha}] &=& {{\dot u}_\alpha}{{\bf e}_0}
-(H {{\delta _\alpha}^\beta}
+{{\sigma _\alpha}^\beta})
{{\bf e}_\beta}
\label{commute1}
\\
\left [ {{\bf e}_\alpha },{{\bf e}_\beta} \right ]  &=&
(2 {a_{[\alpha}}{{\delta _{\beta ]}}^\gamma}
+ {\epsilon _{\alpha \beta \delta }}{n^{\delta \gamma}}){{\bf 
e}_\gamma},
\label{commute2}
\end{eqnarray}
where $n^{\alpha \beta}$ is symmetric, and $\sigma ^{\alpha \beta}$ is
symmetric and trace free.
The scale invariant tetrad variables are defined by 
${\parb_0} \equiv 
{{\bf e}_0}/H$ and ${\parb_\alpha} \equiv {{\bf e}_\alpha}/H$
while scale invariant versions of the other gravitational 
variables are given by
\begin{equation}
\{ {{E_\alpha}^i}, {\Sigma _{\alpha \beta }}, {A^\alpha} ,
{N_{\alpha \beta }} \} \equiv \{ {{e_\alpha}^i} ,
{\sigma _{\alpha \beta }} , {a^\alpha}, {n_{\alpha \beta}} \} /H.
\end{equation}
Note that the relation between the scale invariant tetrad variables 
and
the coordinate derivatives is
\begin{eqnarray}
{\parb_0} &=& {{\cal N}^{-1}} {\partial _t}\label{d0_def}
\\
{\parb_\alpha} &=& {{E_\alpha}^i}{\partial _i},
\end{eqnarray}
where ${\cal N} = NH$ is the scale invariant lapse.  
The matter model is a scalar field $\phi$ with potential $V$  
of the form
\be\label{vdef}
V(\phi) = - V_0 e^{-k \phi},
\ee
where $V_0$ and $k$ are positive constants.
The scale invariant matter variables are given by
\begin{eqnarray}
W &=& {\parb _0} \phi \label{wdef}
\\
{S_\alpha} &=& {\parb_\alpha} \phi
\\
{\bar V} &=& V /{H^2} \label{vbdef}.
\end{eqnarray}
The time coordinate $t$ is chosen so that 
\begin{equation}
{e^{-t}} = 3 H.
\label{timechoice}
\end{equation}
Here we have used the scale invariance of the physical system to make
both $t$ and $H$ dimensionless quantities.
Note that equation (\ref{timechoice}) 
means that the surfaces of constant time are constant 
mean 
curvature surfaces. Note also that the singularity is approached as
$t \to - \infty$.  

\noindent Due to equation (\ref{timechoice}) the scale invariant lapse satisfies 
an elliptic equation
\begin{eqnarray}
&-& {\parb ^\alpha}{\parb _\alpha} {\cal N} + 2 
{A^\alpha} {\parb _\alpha} {\cal N} 
\nonumber
\\
&+& {\cal N} (3 +
{\Sigma _{\alpha \beta}}{\Sigma ^{\alpha \beta}} + {W^2} - {\bar V}) = 
3 \label{Neqn}.
\end{eqnarray}
The gravitational quantities ${{E_\alpha}^i}, \, {A_\alpha}, 
{N^{\alpha \beta}}$ and $\Sigma _{\alpha \beta}$ satisfy the following 
hyperbolic evolution equations
\begin{eqnarray}
{\partial _t} {{E_\alpha}^i} &=& {{E_\alpha}^i} - {\cal N} 
({{E_\alpha}^i}
+ {{\Sigma _\alpha}^\beta}{{E_\beta}^i})
\\
{\partial _t} {A_\alpha} &=& {A_\alpha} + {\textstyle {\frac 1 2}} 
{{\Sigma _\alpha}^\beta}{\parb _\beta}{\cal N} -
{\parb _\alpha}{\cal N} 
\nonumber
\\
&+& {\cal N} \left (
 {\textstyle {\frac 1 2}} {\parb _\beta}{{\Sigma _\alpha}^\beta}
- {A_\alpha} - {{\Sigma _\alpha}^\beta}{A_\beta} \right )  
\\
{\partial _t} {N^{\alpha \beta}} &=& {N^{\alpha \beta}} - 
{\epsilon ^{\gamma \delta ( \alpha}}{{\Sigma _\delta}^{\beta )}}
{\parb _\gamma} {\cal N} 
+ {\cal N} \Bigl ( - {N^{\alpha \beta}}
\nonumber
\\
&+& 2 {{N^{(\alpha}}_\gamma}{\Sigma ^{\beta )\gamma}} - 
{\epsilon ^{\gamma \delta ( \alpha}} {\parb _\gamma} {{\Sigma 
_\delta}^{\beta )}} \Bigr )   
\end{eqnarray}
\begin{eqnarray}
{\partial _t} {\Sigma _{\alpha \beta}} &=& {\Sigma _{\alpha \beta}} 
+ {\parb _{<\alpha}}{\parb _{\beta >}} {\cal N} + {A_{<\alpha}}
{\parb _{\beta >}} {\cal N}
\nonumber
\\
&+& {\epsilon _{\gamma \delta (\alpha}}
{{N_{\beta )}}^\delta}{\parb ^\gamma}{\cal N}
+ {\cal N} \Bigl [ - 3 {\Sigma _{\alpha \beta}} - {\parb _{<\alpha}}
{A_{\beta >}}
\nonumber
\\
&-& 2 {{N_{<\alpha}}^\gamma}{N_{\beta >\gamma}}
+ {{N^\gamma }_\gamma}{N_{< \alpha \beta >}}
\nonumber
\\
&+&  {\epsilon _{\gamma \delta (\alpha}} ( {\parb ^\gamma} {{N_{\beta
)}}^\delta}
- 2 {A^\gamma} {{N_{\beta )}}^\delta}) + {S_{<\alpha}}{S_{\beta >}}
\Bigr ]
\end{eqnarray}
Here parentheses around a pair of indices denote the symmetric part, while
angle brackets denote the symmetric trace-free part.

The equations of motion for the matter variables are as follows:
\begin{eqnarray}
{\partial _t} \phi &=& {\cal N} W
\\
{\partial _t} {S_\alpha} &=& {S_\alpha} + W {\parb _\alpha } {\cal N}
\nonumber
\\
&+& {\cal N} \left [ {\parb _\alpha} W - ({S_\alpha } + {{\Sigma 
_\alpha}^\beta}
{S_\beta})\right ]
\\
{\partial _t} W &=& W + {S^\alpha}{\parb _\alpha} {\cal N} 
\nonumber
\\
&+& {\cal N} 
\left ( {\parb ^\alpha}{S_\alpha} - 3 W - 2 {A^\alpha}{S_\alpha} - 
{\frac {\partial {\bar V}} {\partial \phi}} \right ).
\end{eqnarray}
In addition, the variables are subject to the vanishing of the 
following 
constraint quantities
\begin{eqnarray}
{{({{\cal C}_{\rm com}})}^{\lambda i}} &=& {\epsilon ^{\alpha \beta 
\lambda}}
[ {\parb _\alpha} {{E_\beta}^i} - {A_\alpha} {{E_\beta}^i} ] - 
{N^{\lambda \gamma}}{{E_\gamma}^i}
\label{constraintCOM}
\\
{{({{\cal C}_J})}^\gamma} &=& {\parb _\alpha}{N^{\alpha \gamma}} + 
{\epsilon ^{\alpha \beta \gamma}} {\partial _\alpha}{A_\beta}
- 2 {A_\alpha}{N^{\alpha \gamma}}
\label{constraintJ}
\\
{{({{\cal C}_C})}_\alpha} 
&=& {\parb _\beta}{{\Sigma _\alpha}^\beta}
 - 3 {{\Sigma _\alpha}^\beta}{A_\beta} - {\epsilon _{\alpha \beta 
\gamma}}
{N^{\beta \delta}}{{\Sigma _\delta}^\gamma} 
\nonumber
\\
&-& W {S_\alpha}
\label{constraintC}
\\
\nonumber
{{\cal C}_G} &=& 1 + {\textstyle {\frac 2 3}}{\parb _\alpha}{A^\alpha}
- {A^\alpha}{A_\alpha}
- {\textstyle {\frac 1 6}}{N^{\alpha \beta}}{N_{\alpha \beta}} 
\nonumber
\\
&+& {\textstyle {\frac 1 {12}}}{{({{N^\gamma}_\gamma})}^2} 
- {\textstyle {\frac 1 6}}{\Sigma ^{\alpha \beta}}{\Sigma_{\alpha 
\beta}}  
\nonumber
\\
&-& {\textstyle {\frac 1 6}} {W^2} -  {\textstyle {\frac 1 6}} 
{S^\alpha}
{S_\alpha} -  {\textstyle {\frac 1 3}} {\bar V} 
\label{constraintG}
\\
{{({{\cal C}_S})}_\alpha} &=& {S_\alpha} - {\parb _\alpha}\phi.
\label{constraintS}
\end{eqnarray}

The evolution equations can be freely modified by adding multiples of
the constraints to them.  In particular, for numerical stability we
add a multiple of $ {({{\cal C}_C})}_\alpha$ to the right hand side
of the evolution equation for $A_\alpha$ \cite{dgandcg}.   

The initial data must be chosen so that the constraint equations are 
satisfied, and then the evolution equations will ensure that they 
remain
satisfied.  We find solutions of the constraint equations essentially
the same way as is done for more standard methods of numerical 
relativity:
by using the York method \cite{York}.
That is, we choose certain components of the variables and then 
solve the constraints for the rest.  Our choice is by no means the 
most
general possible one; but it
is general enough that we expect any behavior that emerges from
the evolution of these data to reflect the general behavior of 
singularities
for this type of matter.  This expectation is bolstered by the 
experience of numerical simulations of vacuum 
singularities.  In particular, the greatest restriction on the 
generality of our initial data comes from the fact that we restrict our
studies to deviations in homogeneity along a single spatial direction.  
Our spacetimes thus have two Killing fields.  Nonetheless we choose our
initial data to be the sort that in the vacuum two Killing field 
case\cite{beverlyetal} was sufficiently general that the behavior as
the singularity was approached was the same as that of 
more general initial data
in the case with no symmetries\cite{dgprl}.

In the usual approach to numerical relativity the initial data consists of
the spatial metric and the extrinsic curvature.  The York method then involves
choosing a metric conformally related to the spatial metric and part of a
tensor conformally related to the extrinsic curvature, and then solving for
the conformal factor $\psi$ as well as the rest of the extrinsic curvature.  
Since we are using a tetrad approach, we must also have an initial 
spatial triad consistent with the initial spatial metric.  For simplicity, 
we chose the initial conformal metric to be flat and $(x,y,z)$ to be the
usual cartesian coordinates for that metric, and we choose the spatial triad
to lie along those spatial directions.  Thus, the scale free spatial triad
becomes 
\begin{equation}
{{E_\alpha}^i} = {H^{-1}} {\psi ^{-2}}{{\delta _\alpha}^i}.
\end{equation}
It then follows from equation (\ref{commute2}) that
\begin{eqnarray}
{A_\alpha} &=& - 2 {\psi ^{-1}} {{E_\alpha}^i}{\partial _i}\psi
\\
{N_{\alpha \beta}} &=& 0.
\end{eqnarray}
The shear is essentially the trace-free part of the extrinsic curvature, and
as in the usual approach in numerical relativity, the constraint equations
simplify for a particular rescaling of the trace-free part of the 
extrinsic curvature with the conformal factor.  We therefore introduce
the quantity $Z_{\alpha \beta}$ defined by
\begin{equation}
{\Sigma _{\alpha \beta}} = {\psi ^{-6}} {Z_{\alpha \beta}}.
\end{equation}
Similar considerations apply to the matter variables, leading us to define 
the quantity $Q$ given by
\begin{equation}
W = {\psi ^{-6}} Q.
\end{equation}
Here we will specify $Q, \,\phi$ and a part of $Z_{ik}$  
and solve the constraint equations for the conformal factor 
$\psi$ and the rest of 
$Z_{ik}$.
For convenience of the numerical simulations, we choose 
periodic boundary conditions $0 \le x \le 2 \pi$ with $0$ and $2 \pi$ 
identified where $x$ is the single spatial coordinate that the metric 
and matter variables depend on.  Also identifying the other spatial 
coordinates $y$ and $z$ means that our simulation is of a spacetime
with spatial topology $T^3$.  Since the variables depend only on $x$
and since $x$ is periodically identified, specifying a variable means
giving the coefficients of a Fourier expansion of that variable.  

From equation (\ref{constraintC}) and our ansatz for the 
scale invariant variables we obtain
\begin{equation}
{\partial ^i}{Z_{ik}} = Q {\partial _k} \phi.
\label{divZ}
\end{equation}
In the vacuum case (i.e. for vanishing scalar field) this equation 
simply
becomes the conditions that $Z_{ik}$ is divergence-free, which is in 
turn
simply an algebraic condition on the Fourier coefficients of $Z_{ik}$.  
Note that since $\Sigma _{\alpha \beta}$ must be trace-free, so must
$Z_{ik}$. 
A simple, but still fairly general divergence-free and trace-free 
$Z_{ik}$ is the
following:
%
%
\begin{equation}
{Z_{ik}} = \left ( 
\begin{matrix}\label{zic}
{b_2} & \xi & 0
\\
\xi & a_1\cos x + {b_1} & a_2\cos x
\\
0 & a_2\cos x & -{b_1}-{b_2}-a_1\cos x
\end{matrix}
\right ),
\end{equation}
where $\xi ,\, a_1, \, a_2,\  b_1$ and $b_2$ are constants.
We still keep this divergence-free
part of $Z_{ik}$ but now add to it a piece that has a non-zero 
divergence.  We simply specify the Fourier coefficients of $\phi$ and 
$Q$ via
%
%
\bea\label{Qic}
Q(x,t=0) = \frac{f_1}{H} \cos(m_1 x + d_1) \\
\phi(x,t=0) = f_2 \cos(m_2 x + d_2),\label{phiic}
\eea
where $f_1, m_1, d_1, f_2, m_2$ and $d_2$ are constants.
This turns equation (\ref{divZ}) into an algebraic equation for the 
Fourier 
coefficients of this non-zero 
divergence piece of $Z_{ik}$ which we then solve.
 
Now imposing equation (\ref{constraintG}) our ansatz yields
\begin{eqnarray}
{\partial ^i}{\partial _i} \psi &=& ( {\textstyle {\frac 3 4}} {H^2} - 
{\textstyle {\frac 1 4}} V) {\psi ^5} - {\textstyle {\frac 1 8}}
({\partial ^i}\phi {\partial _i}\phi )\psi
\nonumber
\\
&-& {\textstyle {\frac 1 8}}
({Q^2} + {Z^{ik}}{Z_{ik}}) {H^2} {\psi ^{-7}},
\end{eqnarray} 
which is solved for the conformal factor $\psi$ using the numerical 
methods described below.  

The constraint equations (\ref{constraintCOM}) and 
(\ref{constraintJ}) 
are automatically satisfied by this ansatz.  We then satisfy equation
(\ref{constraintS}) by using the given value of $\phi$ to compute
the initial value of $S_\alpha$.

\subsection{Numerical code}

We discretize the system of equations described in the previous 
section
using second order accurate finite difference methods, with Berger
and Oliger \cite{BO} style adaptive
mesh refinement (AMR)  as provided by the PAMR toolkit 
\cite{pamr_amrd}. 
On a single grid a two-time level, 
Crank-Nicholson(CN)-like discretization scheme is used. 
Standard centered spatial derivative operators are
employed, and in the hyperbolic evolution equations spatial
derivatives (and undifferentiated functions)
are averaged over the two time levels as usual within a CN scheme. 
Kreiss-Oliger dissipation~\cite{KO} is applied, and, although
not necessary for the stability of unigrid evolutions, is important
for reducing unphysical high-frequency solution components
sometimes introduced at refinement boundaries.
The elliptic equations are solved using an FAS (Full-Approximation-
Storage)
multigrid algorithm \cite{brandt}.

The elliptic slicing condition is
incorporated into the Berger and Oliger time-stepping algorithm
using the method described in~\cite{Pretorius:2005ua}. Such
modifications are necessary to take advantage of time-subcycling,
however here we find that we can evolve the system without time-
subcycling
yet keep the time step equal to that of the {\em coarsest}
level in the hierarchy. In other words, with a spatial refinement
ratio of $\rho_{sp}$
a given level $L_i$ in the hierarchy (with level
$L_1$ being the coarsest) 
will have a CFL (Courant-Friedrichs-Lewy) factor
$\lambda_i=\Delta t_i/\Delta x_i = \Delta t_1/\Delta x_i$
equal to $\rho_{sp}{}^{i-1}$ times that of the base level 
$\lambda_1$.
In a typical simulation we use $\lambda_1=0.2$, $\rho_{sp}=2$,
and have run cases where
up to 25 levels of refinement were used, giving 
$\lambda_{25}\approx 3 \times 10^6$.
Our code shows no signs of instability, and exhibits clear second
order convergence.
Though technically the evolution scheme is implicit
due to the CN differencing, at each time step 
the code is able to converge to a solution
within several iterations at most. We surmise that this rather 
atypical 
behavior for the solution of hyperbolic difference equations
is due to the ultralocal nature of the spacetime in the approach to 
the
singularity \cite{beverlyetal,dgprl}.  
This is reflected in the differential equations by the spatial
derivative terms becoming negligible, and hence 
are essentially reduced to a set of ordinary differential equations in 
time,
one at each grid point in the domain.

\section{Results}\label{sec_results}
We have run simulations for a variety of initial
conditions; here we show
results from a single example that demonstrates the generic
behavior : evolution from a highly inhomogeneous,
anisotropic universe with significant curvature at the initial
time to a universe containing distinct volumes of either
smooth, homogeneous $w\gg 1$ matter dominated regions, or 
$w=1$ anisotropic regions. Whenever a $w\gg 1$ region forms it
grows exponentially fast in proper volume relative to $w=1$
regions. 

The particular initial conditions for this example are (\ref{zic}-
\ref{phiic})
\bea
a_1&=&0.70, \ \ a_2=0.10, \ \ \xi=0.01, \nonumber\\
b_1&=&1.80, \ \ b_2=-0.15,  \nonumber\\ 
f_1&=&2.00, \ \ m_1=1, \ \ d_1=-1.7, \nonumber\\
f_2&=&0.15, \ \ m_2=2, \ \ d_2=-1.0, \ \ \nonumber
\eea
and
\be
V_0=0.1, \ \ k=10
\ee
for the scalar field potential parameters (\ref{vdef}).
The same initial data was evolved with several resolutions to confirm
second order convergence; the highest resolution has
$2049$ points on the base level, and up to $12$ additional
levels of 2:1 refinement. 

\begin{figure}
\includegraphics[width=2.75in,angle=-90]{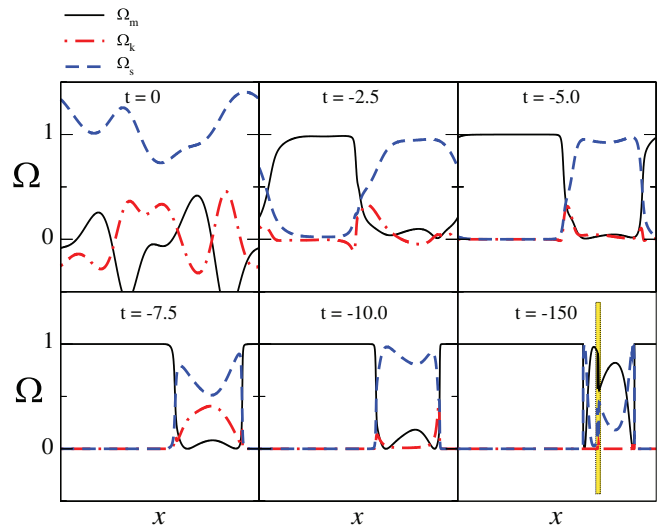}
\caption{$t={\rm const.}$ snapshots of the normalized energy density 
in matter $\Omega_m$
(solid line), curvature $\Omega_k$ (dot-dash line) and
shear $\Omega_s$ (dashed line) 
for $0\le x \le 2 \pi$ at several times during the evolution
of the initial data described in Sec.\ref{sec_results}.
Time runs from left to right along the top row and continues along
the bottom row. 
The shaded slit (dotted outline) in the last panel ($t = -150$)
indicates the range of x shown in the blow-up in Fig.\ref{omega_inset}.
\label{omega_panel}}
\end{figure}

\begin{figure}
\includegraphics[width=2.25in,angle=-90]{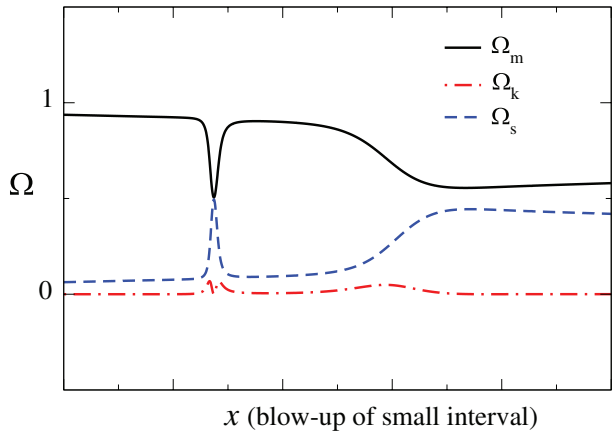}
\caption{Zooming in on one of the spike structures
that formed in the anisotropic region at $t=-150$,
as shown in Fig.\ref{omega_panel}.
\label{omega_inset}}
\end{figure}

It is enlightening to visualize the evolution via the behavior
of the matter ($\Omega_m$), shear ($\Omega_s$) and curvature 
($\Omega_k$) 
contributions to the normalized energy density, defined as
\bea
\Omega_m &\equiv&  
   {\textstyle {\frac 1 6}} {W^2} +  {\textstyle {\frac 1 6}} 
{S^\alpha}
   {S_\alpha} +  {\textstyle {\frac 1 3}} {\bar V} \\
\Omega_s &\equiv&
   {\textstyle {\frac 1 6}}{\Sigma ^{\alpha \beta}}{\Sigma_{\alpha 
\beta}}  \\
\Omega_k &\equiv&
 - {\textstyle {\frac 2 3}}{\parb _\alpha}{A^\alpha}
 + {A^\alpha}{A_\alpha} \nonumber \\
 &\ &+ {\textstyle {\frac 1 6}}{N^{\alpha \beta}}{N_{\alpha \beta}} 
     -  {\textstyle {\frac 1 {12}}}{{({{N^\gamma}_\gamma})}^2},
\eea
where $\Omega_m + \Omega_s + \Omega_k=1$ by (\ref{constraintG}).
Fig.\ref{omega_panel} shows these quantities plotted at select
times during the evolution (which was stopped at $t=-150$). 
Note that all features in the solution are locally
smooth---apparent step functions
in some of the plots are simply due to the large size
of the domain relative to the size of the feature.
As an example, Fig.\ref{omega_inset} shows a zoom-in of the last
panel of Fig.\ref{omega_panel}
about one of the late-time spike structures that formed
in the anisotropic regime.  
\begin{figure}
\includegraphics[width=2.32in,clip,angle=-90]{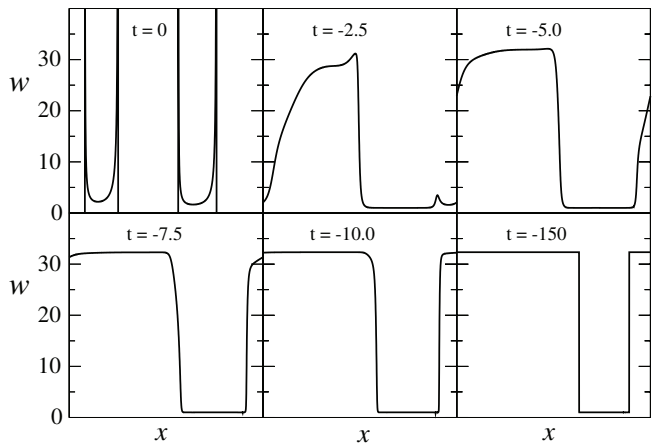}
\caption{The effective equation of state parameter $w$ (\ref{weffdef})
for the simulation described in Sec.\ref{sec_results}, 
for $0 \le x \le 2 \pi$ at the
same times as in Fig.\ref{omega_panel}. At late times 
$w\rightarrow k^2/3-1$ 
in the matter dominated region, and $w\rightarrow 1$ in the
anisotropic region (in this simulation $k=10$ for the 
potential (\ref{vdef})).
\label{w_panel}}
\end{figure}

The effective equation of state parameter $w$ is shown
in Fig.\ref{w_panel}, which takes the following form in
Hubble normalized variables:
\be\label{weffdef}
w=\frac{\frac{1}{2}W^2+\frac{1}{2}S^\alpha S_\alpha-\bar{V}}
       {\frac{1}{2}W^2+\frac{1}{2}S^\alpha S_\alpha+\bar{V}}.
\ee
By comparing Fig.\ref{omega_panel} and Fig.\ref{w_panel} it is
evident that at late times the region that has smoothed
out and become matter dominated coincides with $w \gg 1$,
whereas the anisotropic regime evolves to $w=1$ (and exactly
so to within numerical truncation error). 

The asymptotic behavior of the spacetime in the matter dominated
region appears to coincide with that of a spacetime with an 
isotropic singularity in the sense of Goode and Wainwright\cite{GW}.
We therefore conjecture that in an ekpyrotic phase with an exponential
potential, an open set of initial conditions leads to an isotropic
singularity.
 
We can understand the behavior of the solution in the asymptotic 
matter dominated region by applying the following approximation
which we expect to hold to arbitrary accuracy as the singularity
is approached.  To begin with, we neglect all spatial derivatives 
(such an approximation also holds in the anisotropic regions away from the
isolated spikes). The constraint (\ref{constraintG})
then reduces to
\be\label{approxcG}
\frac{W^2+2\bar{V}}{6} - 1 \approx 0,
\ee
and the slicing condition for ${\cal N}$ (\ref{Neqn})
becomes
\be\label{approxN}
3 {\cal N} \approx \frac{3}{3-\bar{V}},
\ee
where we have used (\ref{approxcG}) to simplify the expression.
We then assume that $\bar{V}$ remains 
finite and non-zero as the singularity is approached. This implies from 
(\ref{vdef},\ref{vbdef},\ref{timechoice}) 
that $\phi$ takes the asymptotic form
\be
\phi(x,t) \approx \phi_0(x) + \frac{2t}{k}
\ee
where $k$ is the constant in the expression for the potential
$V = -{V_0} \exp (-k\phi)$.   
Thus $W$ (\ref{wdef}) tends to
\be
W\approx \frac{2}{k{\cal N}}.
\ee
Combining these relations gives
\bea
W&\approx& k,\\
\bar{V}&\approx& 3-\frac{k^2}{2},\\
{\cal N}&\approx&\frac{2}{k^2},
\eea
and from (\ref{weffdef}) 
\be
w\approx k^2/3 - 1.
\ee
We conjecture that at the singularity the above approximations
become exact, namely
\bea
\lim_{t\to-\infty} W &=& k,\\
\lim_{t\to-\infty} \bar{V}&=& 3-\frac{k^2}{2},\\
\lim_{t\to-\infty} {\cal N}&=&\frac{2}{k^2},\\
\lim_{t\to-\infty} w &=& k^2/3 - 1.
\eea
Our simulations support this
conjecture in that at late times ($t=-150$ in this example) these
quantities have, to within numerical truncation error, reached
their limiting values.
Detailed analysis of the asymptotic dynamics can be carried out 
following the method of~\cite{Lim04,Coley04,Coley05}.

\begin{figure}
\includegraphics[width=2.25in,angle=-90]{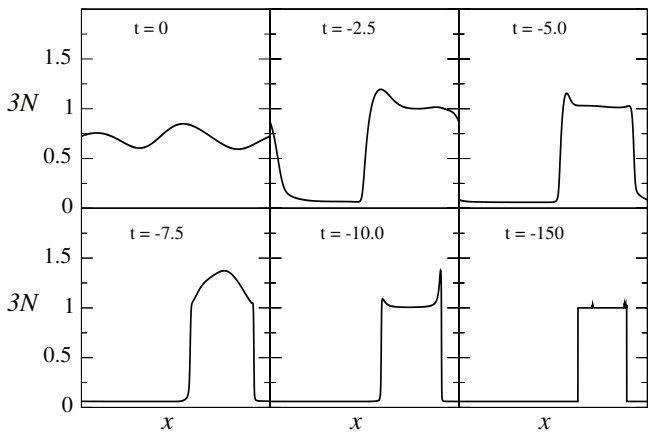}
\caption{3 times the scale invariant lapse ${\cal N}$ for the simulation
described in Sec.\ref{sec_results}, 
for $0 \le x \le 2 \pi$ at the same output times
as Figs.\ref{omega_panel} and \ref{w_panel}.
\label{lapse}}
\end{figure}

Fig.\ref{w_panel} shows that as the singularity
is approached the {\em coordinate} volumes of the $w=1$
vs. $w\gg 1$ regions of the universe are comparable.
However, it turns out that the ratio of the {\em proper}
volume of matter to anisotropic regions grows exponentially
with time. Let $S$ denote the proper spatial volume
element associated with the spatial metric $h_{ij}$ 
of $t={\rm const.}$ slices, i.e., $S=\sqrt{{\rm det}\ h}$.
The fractional change of $S$ with respect to time is
\be
\partial_t{\ln S} = -\frac{1}{2}h_{ij}\partial_t h^{ij},
\ee
which can be written as
\be\label{approxS}
\partial_t{\ln S} = 3{\cal N}.
\ee
Thus the scale invariant lapse ${\cal N}$ is a direct measure of the
rate at which the local proper volume element changes
with time (and recall that $t\rightarrow -\infty$ as the
singularity is approached). Fig.\ref{lapse} shows $3 {\cal N}$
from the simulation at several times. In the asymptotic
regime where spatial gradients are negligible, 
${\cal N}$ approaches a constant (\ref{approxN}), and thus
(\ref{approxS}) can be integrated to give
\bea
S_m &\propto& e^{6t/k^2}, \ \ \ w \gg 1\\
S_v &\propto& e^{t}, \ \ \  w=1 
\eea
where we have used (\ref{approxN}) where $w\gg 1$, and
note that $\bar{V}\approx 0$ when $w=1$.
Thus, at late times the ratio ${\cal R}$ of the proper volume
of matter to anisotropic regions of the universe grows
as
\be
{\cal R}=\frac{\int S_m dx}{\int S_v dx} \propto e^{-t(1-6/k^2)}.
\ee
Thus, as long as $k > \sqrt{6}$ (which is equivalent to $w>1$),
${\cal R}\rightarrow\infty$ as $t\rightarrow-\infty$.

\begin{figure}
\includegraphics[width=3.40in]{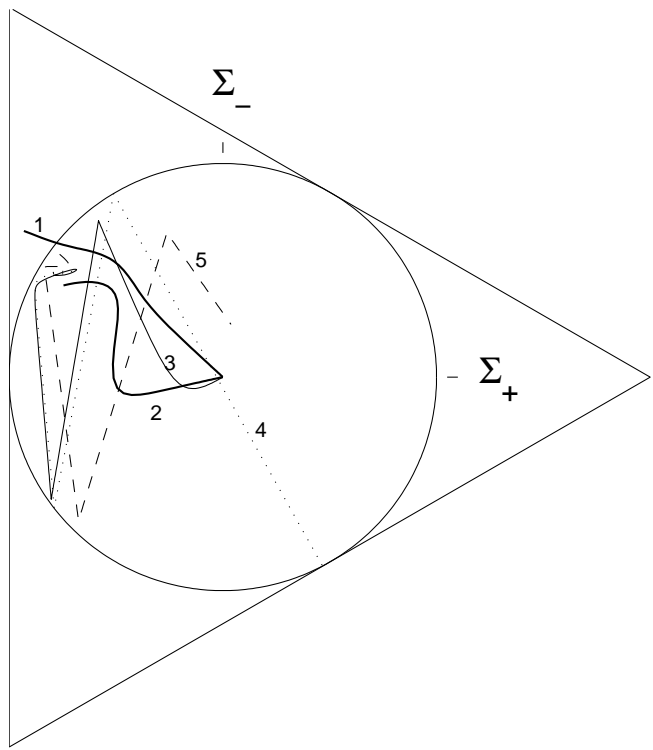}
\caption{The state space orbits for worldlines at
$x=0, 3.0, 3.9, 4.0, 4.4$. Towards the singularity, the first three orbits 
(solid lines)
approach the origin of the plane, indicating isotropization. The 
fourth (dotted) and
the fifth (dashed) do not isotropize. 
\label{trajectories}}
\end{figure}

Fig.\ref{trajectories} shows five state space orbits projected onto 
the
$(\Sigma_+,\Sigma_-)$ plane, where
\be
\Sigma_+ = \frac{1}{2}(\Sigma_{11}+\Sigma_{22}),\quad
\Sigma_- = \frac{1}{2\sqrt{3}}(\Sigma_{11}-\Sigma_{22}).
\ee
The orbits correspond to the evolution along the worldlines at
$x=0, 3.0, 3.9, 4.0, 4.4$.
All five orbits begin in the upper left quadrant, away from the 
origin,
indicating anisotropic initial data.
Towards the singularity, the first three orbits (solid lines)
approach the origin of the plane, indicating isotropization. The 
fourth (dotted) and
the fifth (dashed) do not isotropize.

\section{Conclusions}\label{sec_conclusion}

Our computations provide evidence that the ekpyrotic mechanism for 
smoothing and flattening the universe is robust and powerful, 
comparable qualitatively and quantitatively to the inflationary 
mechanism incorporated in the conventional big bang model.  
This evidence is the behavior as the singularity is approached of a class
of spacetimes that, while not completely general, contain several degrees of
freedom and begin far from FRW.
Both the inflationary and the ekpyrotic mechanism 
require the addition of an energy component that is 
commonly mocked up as a scalar field with potential energy.  
For inflation, the important feature is that, for some initial 
conditions, the scalar field can act like a fluid 
with $w \approx -1$.  It has been shown numerically that, beginning 
from highly non-linear and highly irregular initial conditions, regions 
dominated by the scalar field and with $w \approx -1$ come to 
dominate the volume of the universe\cite{Goldwirth}.
Similarly, we have found 
the regions in which the scalar field acts like a fluid 
with $w >1$ come to dominate exponentially the volume of the 
universe during a contracting phase.   

This result addresses one of the key criticisms raised 
when the ekpyrotic model of the universe was first introduced; 
namely, it was suggested that the model required smooth initial 
conditions \cite{Kallosh:2001ai}.   One of the motivations 
for extending the ekpyrotic picture into a cyclic model 
was to include a period of dark energy domination before the 
ekpyrotic phase began in order to prepare smooth 
conditions  \cite{Steinhardt:2002ih}.  Now, from the results 
here, it is clear that the dark energy epoch is not required for this 
purpose.  Not only does this allow the possibility that the 
dark energy phase lasts only a few e-folds in the cyclic 
picture, as suggested in \cite{Erickson:2006wc}, but it 
also opens the way for more general bouncing cosmologies that 
incorporate the ekpyrotic mechanism but do not cycle.  

With these results in hand, we are now prepared to tackle 
the bounce itself in the case that it is non-singular ($a(t)$ 
shrinks to a non-zero value and then begins to increase).  
For the non-singular bounce, the equation of state must decrease 
from $w>1$ to $w< -1$ for a finite period during which anisotropy 
and inhomogeneity grows.  Our goal is to determine if their growth 
can be kept at a level consistent with observations, establishing 
the viability of these bouncing cosmological models.

\section*{Acknowledgments}
{
This work is supported in part by
the US Department of Energy grant DE-FG02-91ER40671 (PJS),
by NSF grant PHY-0456655 (DG), by the
Alfred P. Sloan Foundation (FP) and NSF PHY-0745779 (FP).
Some computations were run
on the Woodhen cluster at the Princeton Institute for Computational
Science and Engineering (PICSciE).
}


\begin{thebibliography}{1}

\bibitem{Khoury:2001wf}
  J.~Khoury, B.~A.~Ovrut, P.~J.~Steinhardt and N.~Turok,
  Phys.\ Rev.\  D {\bf 64}, 123522 (2001)
  
\bibitem{Erickson:2003zm}
  J.~K.~Erickson, D.~H.~Wesley, P.~J.~Steinhardt and N.~Turok,
  Phys.\ Rev.\  D {\bf 69}, 063514 (2004).
  

\bibitem{Buchbinder:2007ad}
  E.~I.~Buchbinder, J.~Khoury and B.~A.~Ovrut,
  Phys.\ Rev.\  D {\bf 76}, 123503 (2007).
  
  
\bibitem{Steinhardt:2002ih}
  P.~J.~Steinhardt and N.~Turok,
  Science {\bf 296}, 1436 (2002).

\bibitem{kolb_turner_86}
  see for example Sec. 8.6 of
  E.Kolb and M.S.Turner,
  {\em The Early Universe}, 
  Addison-Wesley, Redwood City, CA. (1994)

\bibitem{Goldwirth}
D.~S.~Goldwirth,
Phys. Rev. {\bf D43}, 3204 (1991).



\bibitem{Uggla:2003fp}
  C.~Uggla, H.~van Elst, J.~Wainwright and G.~F.~R.~Ellis,
  Phys.\ Rev.\  D {\bf 68}, 103502 (2003)

\bibitem{Curtis:2005va}
  J.~Curtis and D.~Garfinkle,
  Phys.\ Rev.\  D {\bf 72}, 064003 (2005)

\bibitem{BO} M.J. Berger and J. Oliger,
 {\em J. Comp. Phys. } {\bf 53}, 484 (1984)

\bibitem{Pretorius:2005ua}
  F.~Pretorius and M.~W.~Choptuik,
  J.\ Comput.\ Phys.\  {\bf 218}, 246 (2006)

\bibitem{dgspike}
D.~Garfinkle,
Class.\ Quant.\ Grav.\ {\bf 21}, S219 (2004)

\bibitem{Coley05}
A.~A.~Coley and W.~C.~Lim,
Class.\ Quant.\ Grav.\ {\bf 22}, 3073 (2005)

\bibitem{dgaei}
D.~Garfinkle,
Class.\ Quant.\ Grav.\ {\bf 24}, S295 (2007)

\bibitem{dgandcg}
D.~Garfinkle and C.~Gundlach,
Class.\ Quant.\ Grav.\ {\bf 22}, 2679 (2005)  

\bibitem{York}
J.~W.~York,
Phys.\ Rev.\ Lett.\ {\bf 26}, 1656 (1971) 

\bibitem{beverlyetal}
B.~K.~Berger, D.~Garfinkle, J.~Isenberg, V.~Moncrief, and M.~Weaver,
Mod.\ Phys.\ Lett.\ {\bf A13}, 1565 (1998)

\bibitem{dgprl}
D.~Garfinkle
Phys.\ Rev.\ Lett.\ {\bf 93}, 161101 (2004)  

\bibitem{pamr_amrd} PAMR (Parallel Adaptive Mesh Refinement) and AMRD 
(Adaptive Mesh Refinement Driver)
libraries ({\small {\tt 
http://laplace.physics.ubc.ca/Group/Software.html}})

\bibitem{KO} H. Kreiss and J. Oliger,
{\em Global Atmospheric Research Programme, Publications Series No. 
10.} (1973)

\bibitem{brandt} A. Brandt,
 {\em Math. Comput. } {\bf 31}, 333 (1977)

\bibitem{GW}
S.~W.~Goode and J.~Wainwright,
Class.\ Quant.\ Grav.\ {\bf 2}, 99 (1985)

\bibitem{Lim04}
W.~C.~Lim, H.~van~Elst, C.~Uggla and J.~Wainwright,
Phys.\ Rev.\  D {\bf 69}, 103507 (2004)

\bibitem{Coley04}
A.~A.~Coley, Y.~He and W.~C.~Lim,
Class.\ Quant.\ Grav.\ {\bf 21}, 1311 (2004)

\bibitem{Kallosh:2001ai}
  R.~Kallosh, L.~Kofman and A.~D.~Linde,
  Phys.\ Rev.\  D {\bf 64}, 123523 (2001)

\bibitem{Erickson:2006wc}
  J.~K.~Erickson, S.~Gratton, P.~J.~Steinhardt and N.~Turok,
  Phys.\ Rev.\  D {\bf 75}, 123507 (2007)

\end{thebibliography}
\end{document}